\documentclass[english,superscriptaddress,prl,twocolumn]{revtex4}
\usepackage[toc,page,title,titletoc,header]{appendix}
\usepackage[T1]{fontenc}
\usepackage[latin1]{inputenc}
\usepackage{amsfonts,amssymb}
\usepackage{graphicx}
\usepackage{amsmath}
\usepackage{shapepar}
\usepackage{color}
\usepackage[colorlinks=true,citecolor=blue,linkcolor=blue,anchorcolor=blue,urlcolor=blue]{hyperref}
\hypersetup{linktoc=all}
\makeatletter
\usepackage{babel}
\makeatother

\begin{document}
\title{Theory-Independent Measure of Coherence}

\author{Liang-Liang Sun,   Fei-Lei Xiong,  Sixia Yu\footnote[1]{email: yusixia@ustc.edu.cn}
, Zeng-Bing Chen\footnote[2]{email: zbchen@ustc.edu.cn}
}
\affiliation{Hefei National Laboratory for Physical Sciences at Microscale and Department
of Modern Physics, University of Science and Technology of China, Hefei,
Anhui 230026}
\date{\today{}}

\begin{abstract}

 We provide a theory independent framework to quantify coherence.  In comparison with Bell's theory independent approach to quantum nonlocality, we  characterize a general coherence  phenomenon with statistics arising from sequential measurements of observables that might not be compatible. By introducing a "decohered" state after the sharp measurement of some preferred observable,  we quantify coherence by either the  difference of initial "superposed" state  from the "decohered" state or the measurement outcome deviations when they subjected to further measurements.   Applied to quantum mechanics, the outcome-difference measures yield two novel quantum coherence measures,  one of which  upper-bounds  quantum  interference visibility. In the Bell's scenario, we find a finite gap of coherences between a super non-local model and quantum mechanic and therefore our framework can help to single out quantum mechanics  beyond non-locality.

\end{abstract}


\maketitle

\emph{Introduction}.---Coherence is a fundamentally non-classical feature.  It is the resource of many quantum phenomena that  provide advantages in practical applications over classical resources~\cite{1,2,3,4,5,6,7}.
In quantum mechanics (QM), coherence originates from the superposition principle, which states that a superposition of  valid states forms a new valid state.  Recently, considerable work has been devoted to characterizing and quantifying   quantum  coherence~\cite{Cor, Cd, CE, CR, Cop, Cro1, Cro2,rmp,Bis}.  These coherence measures were either  defined by functions of a density matrix's off-diagonal entries, or  proposed with an analogy with entanglement theory~\cite{Cor,rmp}. These measures then have been   studied on the foundations of quantum resource theory,    leading many results~\cite{rmp}.

 On the other hand,  studying quantum phenomena   in a theory-independent    manner enables one  to find physical ground behind their formulism~\cite{gs,Barr,bel,gpt,ic,wvd,cl}, then yielding a deep understanding of QM.  For example,  Bell's inequalities only involve  necessary  constitutive components  while  not relying  on any specific structures~\cite{bell},  thus  appliable  to any theory~\cite{bell,CHsh,pr}. By the inequalities, it has been found that the no-signalling principle is not enough to specify  quantum correlation~\cite{pr}, therefore triggering   fruitful research for principles constraining QM correlation~\cite{ic,wvd,cl}. It has been found that many features, thought special to QM,  can be studied in a theory-independent manner, $e.g.$, teleportation~\cite{ta,te},  purification~\cite{pur}  and entanglement swapping~\cite{sw}. Coherence is another fundamental QM  feature, and a theory-independent quantification of it would not only provide  intuitive coherence measures  but also provide a new viewpoint to specify   QM. However,  such a framework is missing.

In this Letter, we provide a theory-independent coherence quantification framework. The framework focuses on the  features  attributed to coherence while  not relying on  the specific  structure of a theory.    We  introduce  classical mixture counterpart (CMC) for a  state corresponding to a general physical system (like the local realistic theory in Bell's inequalities). Coherence is characterized by the  difference of ``superposed'' system  from  its CMC, and their outcome deviations when subjected to experiment.   We  then give two kinds of coherence measures, $i.e.$,  ensemble-difference (ED) measure and  measurement-outcome-difference (MOD) measure.  For the MOD measure,  we employ Kolmogorov distance and fidelity distance to quantify the outcome difference, denoted as MOD-K and MOD-F, respectively. We then apply the framework to  QM,  and find that quantum ED measure coincides with  the relative entropy measure,   while both quantum MOD-K and quantum MOD-F are novel QM coherence measures. As quantum phase-sensitive-interference-visibility (PSIV)  is often taken as a coherence measure~\cite{Bis,wd1,wd2,wd3},  we compare quantum MOD-K with  PSIV, find that MOD-K  captures all the coherence carried by a quantum state and upper-bounds the  PSIV.    Finally,  coherence predicted by Popescu and Rohirlich box model (PR-box)~\cite{pr} is studied,  and difference between PR-box coherence and QM coherence is shown.

\emph{Superposition and classical mixture}.---The key observation from the two-slit interference  experiment is that: the photons do not pass through  either upper or lower slit independently, but pass through the two slits simultaneously~\cite{PHY,wd1,Wh}, and then coherent superposition of  pathes is claimed.  Coherence is defined with respect to a preferred observable here denoted by $A$ and not confined to the path. To be more general, $A$ and another observable $A'$ (which  shall be used) are treated as  general discrete valued observables. Note that even through we are talking about observables and measurements, we do not  deal with them in  QM, and  all the mentioned  measurements  are sharp measurements.  For a general  underlying theory, we only attach several necessary constitutive properties~\cite{re,Von}:   for any qualified observable, there should be a corresponding accurate measurement   referred to as the sharp measurement.  For a sharp measurement, we take measurement of observable $A$ for example, each run would yield an  outcome  $a_{j}$  with probability  $p_{j|A}$ then the state of system $\mathcal{S}$  would be prepared in an  ``eigenstate'' of $A$ denoted as~$\mathcal{S}_{j|A}$; an immediately sequential measurement  of $A$ would repeat the outcome $a_{j}$~\cite{sm1,sm2}.

  After performing  sufficient  amount  circles of measurement $A$ on state ensemble  $\mathcal{S}$,  the post-measurement ensemble   would be transferred to a  mixture of  ``eigenstates'': $\overline{\mathcal{S}_{A}}=\sum_{j}p_{j|A}\mathcal{S}_{j|A}$. $\overline{\mathcal{S}_{A}}$ is defined as  the CMC of $\mathcal{S}$ with respect  to preferred observable $A$. The CMC is introduced as a free coherence   state then taken for the comparation with original state, thus the coherence is characterized.

\emph{Ensemble-difference and measurement-outcome-difference coherence measures}.---As the sharp measurement is repeatable,  the preparation of CMC does not change the statistics of the  preferred observable, while destroys the coherence.  Our  first measure quantifies  the change of ensemble state  due to  coherence destruction. We employ measurement entropy~\cite{e} to identify  an state, and it is defined as: $\mathbb{S}(\mathcal{S})=\inf_{A'}\{\sum_{j}-p_{j|A'}\log p_{j|A'}\}$, where $\{p_{j|A'}\}$ denotes statistic from measurement of observable $A'$.  One direct coherence measure with respect to $A$ is defined as,
  \begin{eqnarray}
 C_{E}(\mathcal{S})=\left|\mathbb{S}(\mathcal{S})-\mathbb{S}(\overline{\mathcal{S}_{A}}) \right|\, .
\end{eqnarray}
 ED coherence measure is theory-independent as it does not rely  on specific structure.

 Our second measure is motivated by quantifying the difference between observable ``fringes'' from measurements on  "photons" with  superposed  ``pathes'' and that with  mixed  ``pathes''.   We denote  by $\{p_{i|A'}\}$ and  $\{p^A_{i|A'}\}$ the statistics of measurement  $A'$  on $\mathcal{S}$ and   $\overline{\mathcal{S}_{A}}$, respectively,  where $p^A_{i|A'}=\sum_{j}p_{j|A}p_{i;A'|j;A}$ and $p_{i;A'|j;A}$  denotes probability of obtaining $i$ when measuring $A'$ on `eigenstate' of $A$ with value $j$.  The MOD coherence measure reads
\begin{eqnarray}
 C_M=\sup_{A'}\operatorname{Dis}(\mathbf{p}_{A'},\mathbf{p}^A_{A'})\, .
\end{eqnarray}
  We have many distance measures for probability distribution. In the following, we take the  Kolmogorov distance and fidelity distance~\cite{1}, with corresponding measures  referred to as MOD-K (denoted as $C_{M}^{\rm K}$) and MOD-F (denoted as $ C_M^{\rm F}$), respectively,
\begin{eqnarray}\label{2}
C_{M}^{\rm K}&=&\frac{1}{2}\sup_{A'}\sum_{i}|p_{i|A'}-p^A_{i|A'}|\,,\\
C_M^{\rm F}&=&1-\inf_{A'}\operatorname{F}(\mathbf{p}_{A'}, \mathbf{p}^A_{A'})\,. \label{3}
\end{eqnarray}
Here, the fidelity is defined as
$\operatorname{F}(\mathbf{p}_{A'}, \mathbf{p}^A_{A'})=\sum_{i}\sqrt{p_{i|A'}p^A_{i|A'}}$.
From above definitions, it can be seen that these measures are also theory-independent.

\emph{Application  to  QM}.---A qualified theory-independent measure  should yield a well-defined special measure when applied to a given theory.   We now apply our framework to QM and consider a quantum state's coherence.  In QM, a state of system is represented as a density matrix $\mathcal{S}=\rho$, and the sharp measurement as projective measurement~\cite{sm1,sm2,Von}. Thus CMC of $\rho$   with respect to preferred observable $A$ is given as  $\overline{\mathcal{S}_{A}}\equiv\sum_{i_A}|i_A\rangle\langle i_A|\rho|i_A\rangle\langle i_A|:= \rho_A$, where $|i_A\rangle\langle i_A|$ denotes the $i$-th eigenstate of $A$~\cite{re,Von} and $\rho_A$ denotes the completely decohered state.

In QM, we have  $\mathbb{S}(\rho)=S(\rho):=-\operatorname{Tr}\left(\rho\log\rho\right)$, which is von Neumann entropy~\cite{1,Von,e}.   The ED measure is given as
\begin{eqnarray}
C_{E}(\rho)=S(\rho)-S(\rho_A)=S(\rho\|\rho_A)\,,
\end{eqnarray}
where $\operatorname{Tr}\left(\rho_A\log\rho_A\right)=\operatorname{Tr}\left(\rho\log\rho_A\right) $ has been used.  $S(\rho\|\rho_A)$ is the relative entropy between $\rho$ and $\rho_A$, and it has been studied as a well-defined coherence measure  in~\cite{Cor,rmp}.

We now show that MOD  yields two novel quantum coherence measures. Consider the quantum version of MOD-K, it can  be  written as
\begin{eqnarray}\nonumber
C_{M}^{\rm K}&=&\frac{1}{2}\sup_{A'}\sum_{i}\left|p_{i|A'}-p^A_{i|A'}\right|\\ \nonumber
&=&\frac{1}{2}\sup_{A'}\sum_{i}\left|\operatorname{Tr}(\rho-\rho_{A})\hat{P}_{i|A'}\right|\\
&=&\frac{1}{2}\operatorname{Tr}|\rho-\rho_{A}|\,,
\end{eqnarray}
where  $\{\hat{P}_{i|A'}\}$ are projective operators of $A'$, and $|\rho-\rho_{A}|:=\sqrt{(\rho-\rho_{A})^{\dag}(\rho-\rho_{A})}$.  The reason for the third equality is that: the operator $\rho-\rho_A$ is hermitian,  and its maximum  is obtained when bases of $A'$ are the eigenvectors of $\rho-\rho_A$~\cite{1}. Thus,   and the quantum $C_{A}^{\rm K}$ has a simple formula as the trace distance of $\rho$ from $\rho_A$.

We have quantum version of  MOD-F  as
    \begin{eqnarray}\nonumber
C_M^{\rm F}&=&1-\inf_{A'}\operatorname{F}(\mathbf{p}_{A'}, \mathbf{p}^A_{A'})\\
        &=&1-\operatorname{F}_{q}(\rho, \rho_A)\,,
\end{eqnarray}
 where $\operatorname{F}_{q}(\rho, \rho_A)=\operatorname{Tr}\sqrt{\sqrt{\rho}\rho_A \sqrt{\rho}}$ is quantum fidelity between the state $\rho$ and $\rho_A$. In QM, $\inf_{A'}F(\mathbf{p}_{A'}, \mathbf{p}^A_{A'})=F_{q}(\rho, \rho_A)$, which also is a simple formula~\cite{1}.

 Recently, a general quantum resource measure was proposed as a distance of state $\rho$ from its resource-destroyed state $\mathfrak{D}(\rho)$: $\mathfrak{D}(\rho)=D(\rho, \Gamma(\rho))$~\cite{d}. Though,  our quantum coherence measures  are not derived  from a quantum resource theory but in a theory independent manner,  interestingly, in QM, we find that all $C_{E}$, $C_{M}^{\rm K}$ and $C_M^{\rm F}$ formally coincide with $\mathfrak{D}$. Different from the measure $\Gamma(\rho)$, our measures are applicable to a general theory.   We shall apply it  to PR-box model, while $\mathfrak{D}(\rho)$ is confined to QM.

As quantum $C_{M}^{\rm K}$ and  $C_M^{\rm F}$  are novel measures, let us now examine them  with the requirements proposed  in  Ref.~\cite{Cor,rmp,T}, which has been taken  as  qualifications for a well-defined quantum coherence measure:
\begin{enumerate}
  \item $C(\rho) \geq 0$  for all quantum states, and $C(\rho)=0$ if and only if $\rho$ are incoherent states.
  \item $C(\rho) \geq  C(\Lambda(\rho))$ if $\Lambda$ is incoherent operation.
   \item  $C(p_{1}\rho_{1}\oplus p_{2}\rho_{2})=p_{1}C(\rho_{1})+p_{2}C(\rho_{2})$ for Bloch diagonal states $\rho$ in the incoherent basis.
\end{enumerate}
 As the fidelity shares many similar properties with the trace distance~\cite{1}, we only need to examine  the quantum MOD-K with the requirements, and the results are applicable to quantum MOD-F.
Quantum MOD-K definitely satisfies the first requirement. We shall prove the  requirement 2 under dephasing-covariant-incoherent operation. We need to prove
 \begin{eqnarray}\label{M}
\operatorname{Tr}\left|\Lambda(\rho)-\Delta(\Lambda(\rho))\right| \leq \operatorname{Tr}\left|\rho-\Delta(\rho)\right|\,,
\end{eqnarray}
where  $\Delta$  denotes a dephasing map $\Delta(\rho)=\sum_{i}|i\rangle\langle i|\rho|i\rangle\langle i|$, and  it  maps  a state to its CMC. $\Lambda$ denotes a dephasing-covariant-incoherent operation (Chitamber and Gour have made a critical examation on various incoherent operations and showed their  containment relationships~\cite{Cop})
 which  commutes with $\Delta$ by definition,   then $\operatorname{Tr}(|\Lambda(\rho)-\Delta(\Lambda(\rho))|)=\operatorname{Tr}(|\Lambda(\rho)-\Lambda(\Delta(\rho))|)$. We take the trace norm convexity theorem~\cite{1}: the trace distance of  arbitrary states $\rho_{1}$ and  $\rho_{2}$ is contractive under any tracing-preserving quantum operations $\xi$
, $\operatorname{Tr}|\xi(\rho_{1})-\xi(\rho_{2})|\leq \operatorname{Tr}|\rho_{1}-\rho_{2}|$.
 As   $\Lambda\subset\xi$~\cite{1,Cor},    we  have,
    \begin{eqnarray}\nonumber
\operatorname{Tr}|\Lambda(\rho)-\Delta(\Lambda(\rho))|&=&\operatorname{Tr}|\Lambda(\rho)-\Lambda(\Delta(\rho))| \\
&\leq &\operatorname{Tr}|\rho-\Delta(\rho)|\,.
\end{eqnarray}
As for  the requirement 3, it is obvious that
 \begin{eqnarray}\nonumber
&&\operatorname{Tr}\left|p_{1}\rho_{1}\oplus p_{2}\rho_{2}-p_{1}\rho_{1A}\oplus p_{2}\rho_{2A}\right|\\
&&\quad =p_{1}\operatorname{Tr}|\rho_{1}-\rho_{1A}|+ p_{2}\operatorname{Tr}|\rho_{2}-\rho_{2A}|\,.
\end{eqnarray}
Thus  quantum MOD-K  satisfies all  three requirements under dephasing-covariant-incoherent operation.  Although the  proof  under  maximal incoherent operation is still missing, quantum  MOD-K is physically well-motivated. We shall show that the  quantum MOD-K upper-bounds a commonly used interference measure.

 In the two-slit interference experiments, PSIV has been used as a  quantifier of coherence  between two paths~\cite{wd1,wd2,wd3,wds},  which is defined by   the sensitivity of measurement  outcome to  phase shift.   In the schemes ,  an input state $\rho$  is distributed into two ways based on a $\sigma_{z}$--beam splitter, then  undergoes a path-dependent phase shift $ U^{\dag}_{\phi}\rho U_{\phi}$, where $ U_{\phi}=\exp(-\operatorname{i}{\phi}\sigma_{z}/2)$, and merges to a final measurement.  The frequencies of  outcome $+1$ is $\operatorname{I}$,  and the PSIV is commonly defined as~\cite{9}
    \begin{eqnarray}\label{V}
V=\frac{\operatorname{I}_{\max}-\operatorname{I}_{\min}}{\operatorname{I}_{\max}+\operatorname{I}_{\min}}\,.
\end{eqnarray}
where $\operatorname{I}_{\min}=\inf_{\phi} \operatorname{Tr}\left( U^{\dag}_{\phi}\rho U_{\phi}\hat{P}_{+}\right)$, $\operatorname{I}_{\max}=\sup_{\phi} \operatorname{Tr}\left( U^{\dag}_{\phi}\rho U_{\phi}\hat{P}_{+}\right)$, and $\hat{P}_{+}$  is the projector of the final measurement on the direction $(|0\rangle+|1\rangle)/\sqrt{2}$.
 By  trivial calculus, we find that the visibility  is exactly the same as  quantum MOD-K with preferred observable $\sigma_{z}$ up to a constant 2, namely $V=2C_{M}^{\rm K}$.

 With the above observation, we now explore a general relation between PSIV and quantum MOD-K. Consider a general measurement where state $\rho$ is distributed to $d$ pathes based on a "$A$--beam splitter", and each one undergoes a phase shift before merging to a measurement $A'$.   Taking all the frequencies  into consideration,  we generalize Eq.~(\ref{V}) as  $V_{d}=\frac{1}{2}\sup_{\Phi,\Phi'}\sum_{i}|p_{i|\Phi,A'}-p_{i|\Phi',A'}|$,  where $p_{i|\Phi,A'}=\operatorname{Tr}(U^{\dag}_{\Phi}\rho U_{\Phi}\hat{P}_{i|A'})$,   $U_{\Phi}$ denotes  phase shift and  $\rho_{\Phi}=U^{\dag}_{\Phi}\rho U_{\Phi}$.  Quantum MOD-K  is found to upper-bound  the visibility
    \begin{eqnarray}\nonumber
V&&=\frac{1}{2}\sup_{\Phi,\Phi'}\sum_{i}\left|p_{i|\Phi,A'}-p_{i|\Phi',A'}\right|\\ \nonumber
&&\leq\frac{1}{2}\sup_{\Phi,\Phi', A'}\sum_{i}\left|p_{i|\Phi,A'}-p_{i|\Phi',A'}\right|\\  \nonumber
&&=\frac{1}{2}\sup_{\Phi,\Phi'}\operatorname{Tr}\left|\rho_{\Phi}-\rho_{\Phi'}\right|\\  \label {09}
&&\leq\frac{1}{2}\sup_{\Phi, \Phi'}(\operatorname{Tr}\left|\rho_{\Phi}-\rho_{\Phi,A}\right|+\operatorname{Tr}|\rho_{\Phi'}-\rho_{\Phi,A}|)\\  \label {00}
&& =2C_{M}^{\rm K}\,.
\end{eqnarray}
 Equation~(\ref{09}) usually does not saturate,  because the diagonal terms of  $\rho_{\Phi}$ and $\rho_{\Phi'}$ are equal, while their anti-diagonal terms  cancel each other out. Therefore, the coherence of $\rho$ can not be captured by PSIV, while all are captured by the quantum MOD-K.

\emph{Application to PR-box}.---Bell's inequalities triggered  fruitful results on  quantum correlation~\cite{sm1,sm2,MI1}.  The  inequalities consider two space separated observers, Alice and Bob. They share a correlated system and randomly perform measurements $X,Y\in \{0,1\}$ with outcomes $x,y\in\{0,1\}$. The inequalities expose the discrepancies between  general theories.   The model is defined as  $X\oplus Y=x\cdot y$, then we have
\begin{eqnarray}\label{x}
y=X\cdot Y\oplus x \, .
\end{eqnarray}
   For each of Bob's settings and outcomes,   he prepares a state $\omega_{y|Y}$ on Alice's side.   By the definition of the model, any measurement on any Alice's state   would yield outcome in a deterministic way.
In QM, when measured state is an eigenstate of the desired observable, the measurement would yield outcomes in a deterministic way, and the state would not be disturbed. We now show that  PR-box model contradicts with this feature.  If PR-box model shares this feature: measured state would be not disturbed if the outcome is yielded deterministic.  Taking $\omega_{0|0}$ for example, when measurement is $X=0$, then outcome would be deterministic $x=0$. By the assumption of no disturbance, $x=0$ would be obtained for a sequential measurement of $X=1$. Subjecting  $\{X=0, x=0\}$, $\{X=1,x=0\}$ into Eq.~(\ref{x}), one can derive the setting and outcome on Bob's side as $\{Y=1,y=0\}$. Thus if the assumption is applicable to  the  model,  no-instantaneous messaging principle is  violated. To guarantee the principle,  there must be coherence.

 We now apply MOD-K to  the model and consider the coherence of $\omega_{0|Y}$ with respect  to observable  $X=0$.  The state $\omega_{0|Y}$  would give result $x=XY$  deterministic.  We have its coherence
 \begin{eqnarray}\nonumber
C_{M}^{\rm K}(\omega_{0|Y})&\geq&\frac{1}{2}\sup_{X'=0,1}\sum_{i}\left|p_{i|X'}-p^0_{i|X'}\right|\\
&=& \frac{1}{2}\sum_{i}\left| \delta_{Y,i}-p_{i;1|0;0}\right|\,,
\end{eqnarray}
Given the repeatable feature of a sharp measurement,  the maximum is obtained when $X'=1$ is taken, and $p_{i;1|0;0}$ denotes probability of obtaining $i$ when measuring $X=1$ on the state prepared by the measurement $X=0$ with outcome $0$.
As a result, we have 
 \begin{eqnarray}\label{pr}
C_{M}^{\rm K}(\omega_{0|1})+C_{M}^{\rm K}(\omega_{0|0})\geq 1 ,
\end{eqnarray}
for PR box while in QM we have
 \begin{eqnarray}
C_{M}^{\rm K}(\omega_{0|1})+C_{M}^{\rm K}(\omega_{0|0})\le \frac{1}{2}(\Delta_{\omega_{0|1}} A + \Delta_{\omega_{0|0}} A)=0.
\end{eqnarray}
Here, the inequality is because the following upper bound  $C_{M}^{\rm K}= |\langle0_{A}|\rho|1_{A}\rangle|\le \sqrt{p_{0|A}p_{1|A}}=\frac{1}{2}\Delta A$ for the coherence in an arbitrary qubit state.  Thus if the two quantum states yield the same statistics with the $\omega_{0|1}$ and $\omega_{0|0}$ when subjected to reference measurement, their sum of the coherence  must equal to $0$. However, by Eq.~(\ref{pr}), the lower bound of the sum is $1$. A finite gap between the amounts  of their  coherence is shown.

\emph{Conclusion}.---We have  provided a theory-independent coherence-quantification framework.  Our framework gives two novel quantum coherence measures when applied to QM, and has been used to study the coherence of PR-box and specify QM from the model based on the amounts of their coherence. As theory-independent measure of non-locality, Bell's inequalities have triggered fruitful research for  principles constraining QM,  this coherence quantification framework would pave a novel method to specify QM on the other hand.

\emph{Acknowledgement}.---This work has been supported by the Chinese Academy of Sciences, the National Natural Science Foundation of China under Grant No. 61125502, and
the National Fundamental Research Program under Grant No. 2011CB921300.

\end{document}